\documentclass[preprint,12pt]{elsarticle}

\usepackage{amssymb}
\usepackage{amsmath}
\usepackage{color}
\usepackage{booktabs}
\usepackage{colortbl}
\usepackage{graphicx}
\usepackage{mathcomp}
\usepackage{textcomp}

\journal{High Energy Density Physics}

\begin{document}

\begin{frontmatter}



\title{Enhanced plasma heating via interaction with high-contrast laser and cone-shaped target}


\author[label1,label2]{Yuga Karaki} 
\author[label3]{Yoshitaka Mori}
\author[label4,label2]{Eigo Ebisawa}
\author[label5]{Yuichi Inubushi}
\author[label6]{Sadaoki Kojima}
\author[label2]{Kohei Yamanoi}
\author[label4,label2]{Yuki Abe}
\author[label1,label2]{Takumi Tsuido}
\author[label1,label2]{Hiroki Matsubara}
\author[label1,label2]{Rinya Akematsu}
\author[label1,label2]{Ryo Omura}
\author[label2]{Ryunosuke Takizawa}
\author[label2]{King Fai Farley Law}
\author[label7]{Eisuke Miura}
\author[label2]{Yasunobu Arikawa}
\author[label2]{Keisuke Shigemori}
\author[label2,label8]{Akifumi Iwamoto}
\author[label3]{Katsuhiro Ishii}
\author[label3]{Ryohei Hanayama}
\author[label3]{Yoneyoshi Kitagawa}
\author[label9]{Hiroshi Sawada}
\author[label2]{Takayoshi Sano}
\author[label2]{Natsumi Iwata}
\author[label2]{Yasuhiko Sentoku}
\author[label10,label2]{Atsushi Sunahara}
\author[label11,label2]{Tomoyuki Johzaki}
\author[label8]{Kenichi Nagaoka}
\author[label2,label8]{Shinsuke Fujioka}

\affiliation[label1]{organization={Graduate School of Science, The University of Osaka},
            addressline={1-1 Machikaneyama-cho},
            city={Toyonaka},
            postcode={560-0043}, 
            state={Osaka},
            country={Japan}}
\affiliation[label2]{organization={Institute of Laser Engineering, The University of Osaka},
            addressline={2-6 Yamada-Oka}, 
            city={Suita},
            postcode={565-0871}, 
            state={Osaka},
            country={Japan}}
\affiliation[label3]{organization={Graduate School for the creation of New Photonics Industries},
            addressline={1955-1 Kurenatsu-cho, Chuo-ku}, 
            city={Hamamatsu},
            postcode={431-1202}, 
            state={Shizuoka},
            country={Japan}}
\affiliation[label4]{organization={Graduate School of Engineering, The University of Osaka},
            addressline={2-6 Yamada-Oka}, 
            city={Suita},
            postcode={565-0871}, 
            state={Osaka},
            country={Japan}}
\affiliation[label5]{organization={Japan Synchrotron Radiation Research Institute},
            addressline={1-1-1 Koto}, 
            city={Sayo-gun Sayo-cho},
            postcode={679-5198}, 
            state={Hyogo},
            country={Japan}}
\affiliation[label6]{organization={Kansai Photon Science Institute, National Institutes for Quantum and Radiological Science and Technology},
            addressline={8-1-7 Umemidai}, 
            city={Kizugawa},
            postcode={619-0215}, 
            state={Kyoto},
            country={Japan}}
\affiliation[label7]{organization={National Institute of Advanced Industrial Science and Technology},
            addressline={1-1-1 Umezono}, 
            city={Tsukuba},
            postcode={305-8560}, 
            state={Ibaraki},
            country={Japan}}
\affiliation[label8]{organization={National Institute for Fusion Science},
            addressline={322-6 Oroshi-cho}, 
            city={Toki},
            postcode={509-5292}, 
            state={Gifu},
            country={Japan}}
\affiliation[label9]{organization={Department of Physics, University of Nevada, Reno},
            addressline={1664 N Virginia St}, 
            city={Reno},
            postcode={89557}, 
            state={NV},
            country={USA}}
\affiliation[label10]{organization={Center for Materials Under eXtreme Environments (CMUXE), School of Nuclear Engineering, Purdue University},
            addressline={500 Central Drive}, 
            city={West Lafayette},
            postcode={47907}, 
            state={IN},
            country={USA}}
\affiliation[label11]{organization={Graduate School of Advanced Science and Engineering, Hiroshima University},
            addressline={1-3-2 Kagamiyama}, 
            city={Higashi-Hiroshima},
            postcode={739-8511}, 
            state={Hiroshima},
            country={Japan}}

\begin{abstract}
We investigated plasma heating enhancement using a high-intensity, high-contrast laser and a cone-attached target.  
Fast electron spectra and X-ray emission were measured with an electron spectrometer and a Bragg crystal spectrometer.  
The results were analyzed using PrismSPECT simulations with a two-component electron distribution model and empirical scaling laws.  
X-ray pinhole images showed that the cone effectively focused multi-spot laser light near its tip, enhancing local emission.  
While high-contrast laser irradiation reduced the fast electron slope temperature for flat targets, the use of a cone increased it by over threefold, corresponding to a fourfold rise in laser intensity.  
X-ray spectral analysis indicated an electron temperature of ~9~keV for the cone case, 17.5 times higher than that with a low-contrast laser.  
These findings demonstrate that combining high-contrast laser irradiation with cone-target geometry significantly improves laser energy coupling and plasma heating efficiency.
\end{abstract}




\end{frontmatter}



\section{Introduction}
\label{sec1}

The generation of high-energy-density states using short-pulse, high-intensity lasers plays a crucial role in the broad field of high-intensity laser applications. Laboratory-scale experiments that simulate extreme astrophysical environments have significantly advanced our understanding of microscopic physical phenomena that are  difficult to be directly observed~\cite{mourou1998ultrahigh,remington2000review,Law2020a, Sakawa2024}. Experimental validation using high-power lasers is also essential for constructing models of the equation of state and opacity in stellar interiors and astoronomical objects~\cite{da1997absolute,fujioka2005opacity,lefebvre1995transparency,Fujioka2009a}.

In inertial fusion energy research, significant efforts are underway to achieve energy gains of approximately 100 times the laser input by compressing fuel plasma to extremely high temperatures and densities, thereby initiating ignition and burn~\cite{lindl1995development}.
Among these approaches, the fast ignition scheme involves irradiating a highly compressed fuel with a short-pulse, high-intensity laser~\cite{tabak1994_Fast_ignition, kodama2001_nature, theobald2011initial, theobald2014time, fujioka2016fast, Sakata2017, Matsuo2019a, PhysRevResearch.7.023081}. This process forms a localized ignition region and triggers fusion burn. 
Achieving efficient heating of the compressed plasma using short-pulse, high-intensity lasers remains a key challenge.

Short-pulse, high-intensity lasers can accelerate electrons through laser-plasma interaction, generating fast electron beams~\cite{pukhov1999particle}. 
These fast electrons heat the plasma by transfering energy to background electrons, atoms and ions through collisions with those species.
The efficiency of plasma heating is strongly influenced by the contrast of the laser pulse, this influence is a research topic that has attracted extensive research attention~\cite{cai2010prepulse, fujioka2016fast, rosmej2018generation}.

Short-laser pulses typically include low-intensity precursor pulses that arrive ahead of the main pulse. The ratio of the main pulse to the precursor is referred to as the pulse contrast.
When the precursor pulse reaches the target, it generates pre-plasma before the main pulse arrives, which broadens the laser-plasma interaction region~\cite{paradkar2010numerical, peebles2017investigation, yogo2017boosting, kojima2019electromagnetic}.
As a result, more high-energy electrons, which are less efficient at heating plasma, are generated, thus reducing overall heating efficiency~\cite{cai2010prepulse}.
High-contrast laser pulses suppress pre-plasma formation, optimize the energy distribution of fast electrons, and thereby enhance plasma heating efficiency.

In addition to improving laser contrast, introducing cone structures into the target design has been proposed to further enhance heating efficiency. The cone serves as a guiding structure that allows the laser to travels to the high-density region without being significantly absorbed or deflected in low-density and long scale corona plasma surunding the high-density region~\cite{kodama2001_nature}. 
Moreover, fast electron beams generated at the cone tip are spontaneously collimated by the cone structure, reducing their divergence and improving energy coupling~\cite{PhysRevResearch.7.023081}.

In this study, we show that the electron temperature of the plasma increases significantly when a high-contrast laser is applied to a cone-attached target.
This paper is organized as follows.  
Section~2 describes the diagnostic instruments and target configuration used in the experiment.  
Section~3 discusses the enhancement of laser intensity focusing due to the high-contrast laser and cone structure.  
In Section~4, we present the method for estimating plasma electron temperature based on X-ray spectroscopy.  
Finally, Section~5 summarizes the conclusions.

\section{Experimental setup}
\label{sec2}

\subsection{LFEX laser}
\label{subsec2-1}

The LFEX laser at the Institute of Laser Engineering, The University of Osaka, was used to irradiate the target and generate fast electrons~\cite{miyanaga200610}. The LFEX system consists of four laser beams. In this experiment, two of the four beams were used for each shot.
The laser wavelength was 1.053~{\textmu}m, and the pulse contrast exceeded $10^9$ at 2.5~ns before the main pulse~\cite{fujioka2016fast}. The laser energy and pulse duration for each shot are summarized in Table~\ref{tab:condition_table}.

For the shot directly irradiated onto the target (L5146), the uncertainty in laser energy was estimated based on the standard error of the transmission efficiency of the LFEX pulse compressor.
For the shots utilizing a plasma mirror (PM) (L5145 and L5147), the energy uncertainty was evaluated by considering both the variation in PM reflectivity~\cite{arikawa2016_PM} and the transmission efficiency of the pulse compressor.

\begin{table}[ht]
  \centering
  \begin{tabular}{c c c c}
    \toprule
    LFEX shot number & L5145 & L5146 & L5147 \\
    \midrule
    Laser energy [J]        & $137.6 \pm 14.3$ & $161.4 \pm 3.4$  & $123.5 \pm 12.6$ \\
    Pulse duration [ps]     & $1.5 \pm 0.02$   & $1.7 \pm 0.08$   & $1.6 \pm 0.02$   \\
    \bottomrule
  \end{tabular}
  \caption{Laser parameters for each LFEX shot.}
  \label{tab:condition_table}
\end{table}

\subsection{Plasma mirror (PM)}
\label{subsec2-2}

In this study, a plasma mirror (PM) was employed to improve the pulse contrast of the laser system~\cite{kapteyn1991_PM, gold1994_PM, Bdromey2004_PM, doumy2004_PM}. 
The PM used in the experiment was a transparent plano-concave optical element with a diameter of 50.8~mm, featuring a flat surface on one side and a concave surface on the other. 
The concave surface had a radius of curvature of 203.2~mm, corresponding to a focal length of 101.6~mm. 
Both surfaces were coated with anti-reflection (AR) coatings optimized for the LFEX laser wavelength of 1.053~{\textmu}m, achieving an average transmittance of 99.8~\%.
The LFEX laser was initially focused approximately 4~mm below the target before reaching the PM. 
To achieve focus on the target surface, the beam was directed onto the concave side of the PM.

Precursor pulses with intensities below the plasma formation threshold of the glass-based PM transmit through the mirror without generating plasma.
This threshold is approximately 10~J/cm\textsuperscript{2} for a pulse duration of 1.7~ps~\cite{doumy2004_PM}.
When the intensity of the main pulse exceeds this threshold just before its arrival, a plasma is formed on the mirror surface. 
This plasma reflects the main pulse and directs it toward the target.
The pulse contrast is significantly improved by PM as only the high-intensity main pulse is reflected by the PM while the pre-pulses pass through the PM.
The reflectivity of the PM for the main pulse is approximately 45~\%~\cite{arikawa2016_PM}. 
The use of a PM can improve the pulse contrast by about two orders of magnitude.

In this study, the fluence on the PM surface was calculated, and its fluence-dependent reflectivity was taken into account in the analysis of the experimental results.

\subsection{Electron spectrometer}
\label{subsec2-3}

The energy distribution of fast electrons, which were accelerated by an intense laser and emitted into vacuum from the target and surrounding plasma, was measured using an electron spectrometer~\cite{tanaka2005calibration, chen2008high}.
The electron spectrometer mainly consists of three components: a collimator with a diameter of $1000~\mathrm{\mu m}$, a dipole magnet with a width of 40~mm and a length of 249~mm operating at a magnetic field strength of 0.3~T, and an imaging plate (IP) for electron detection~\cite{meadowcroft2008_IP, ozaki2014electron}.

The collimated fast electron beam is deflected by the magnetic field according to its energy. 
By analyzing the deflection pattern recorded on the IP, the energy distribution of the fast electrons can be determined. 
The spatial distribution of the magnetic field and the electron sensitivity of the IP were calibrated in advance.

\subsection{X-ray spectrometer}
\label{subsec2-4}

The resonant X-ray spectrum from titanium was measured using a spectrometer equipped with a Bragg crystal. 
The Bragg crystal used in the setup was rubidium acid phthalate (RbAP), which has a lattice spacing of $2d = 26.120$~\AA{}.
The distance from the center of the plasma to the center of the crystal was 250~mm, and the distance from the crystal center to the detector’s sensitive surface was also 250~mm.
A 150~\textmu m-thick beryllium (Be) filter was placed in front of the crystal to protect it from debris generated during laser irradiation.

When converting the recorded signal to photon counts, we accounted for the transmittance of the Be filter, the reflectance of the RbAP crystal, and the solid angle correction. 
The reflectance of the RbAP crystal was taken from the data reported by P. Seidl \textit{et al.}~\cite{seidl1977diffraction}.
The spectral resolution is determined by the source size, the rocking curve of the RbAP crystal, and the pixel size of the imaging plate. 
In this study, the full width at half maximum (FWHM) of spectral resolution was estimated to be 15~eV.

\subsection{X-ray pinhole camera}
\label{subsec2-5}

X-ray emission from the plasma was recorded using an X-ray pinhole camera to observe the laser irradiation position and focal spot size, 
The pinhole had a diameter of 25~{\textmu}m, and its magnification was 9.8. IP was used as the detector.
The X-ray pinhole camera was positioned at an angle of 65.9\textdegree{} relative to the optical axis of the LFEX laser.

\subsection{Target}
\label{subsec2-6}

\begin{figure}[htbp]
  \centering
  \includegraphics[width=\linewidth]{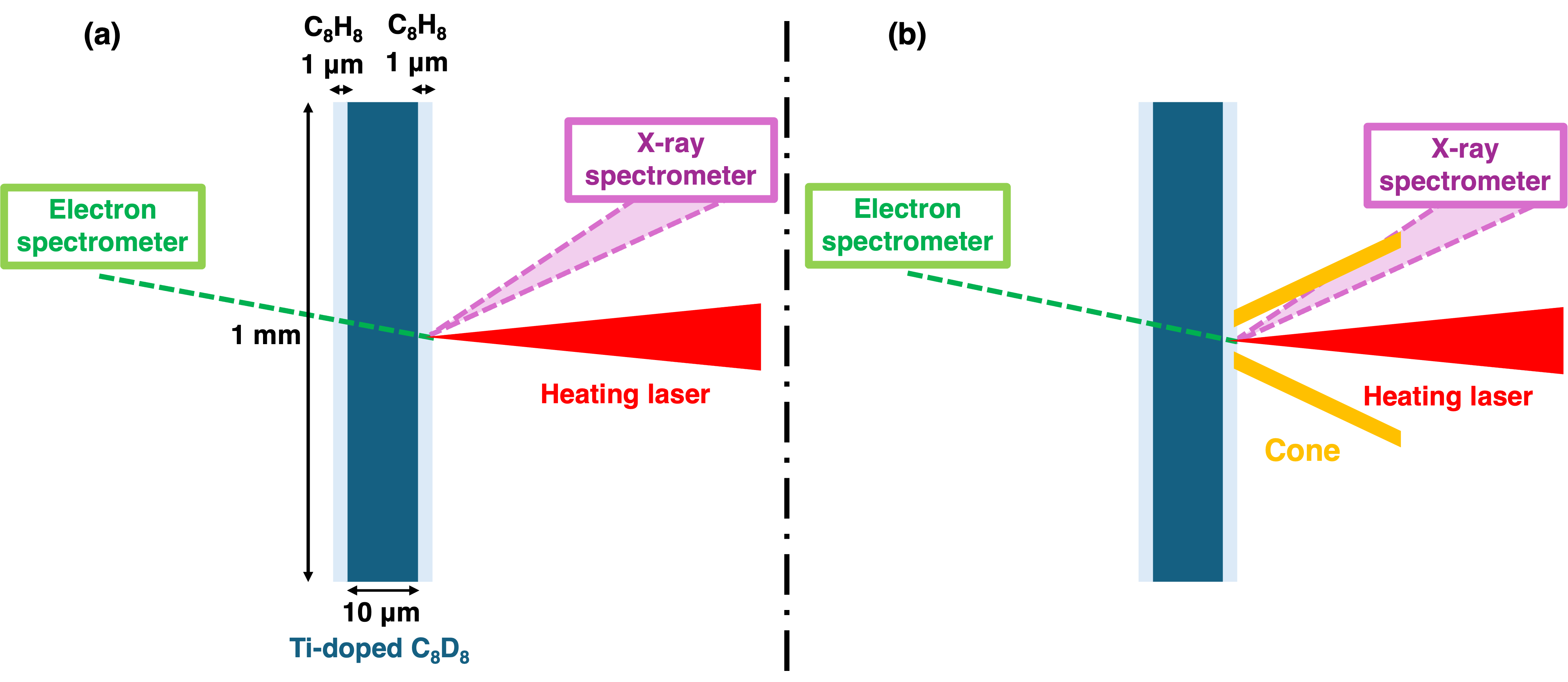}
  \caption{
    (a) Schematic of the planar target composed of deuterated polystyrene ($\mathrm{C_8D_8}$) with dimensions of $1~\mathrm{mm} \times 1~\mathrm{mm} \times 10~\mu \mathrm{m}$. The nano-powders of titanium oxide is dispersed in the deuterated polystyrene with the concentration of 1.3~at\% as a tracer for X-ray spectroscopy. A $1~\mu \mathrm{m}$-thick undoped polystyrene ($\mathrm{C_8H_8}$) layer is placed on the surface to prevent direct laser interaction with the titanium-doped layer.
    (b) Target structure with a gold cone attached. The cone has a thickness of $7 \mu \mathrm{m}$, an opening angle of $45^\circ$, a tip aperture of $100~\mu \mathrm{m}$, and a length of $300~\mu \mathrm{m}$.
  }
  \label{fig:target}
\end{figure}

Figure~\ref{fig:target} shows the structure of the thin-film target used to study heating effects induced by the interaction between the high-contrast laser and the cone, along with the arrangement of diagnostic instruments.

The target consists of deuterated polystyrene ($\mathrm{C_8D_8}$) with lateral dimensions of $1~\mathrm{mm} \times 1~\mathrm{mm}$ and a thickness of $10~\mu \mathrm{m}$. The nano-powders of titanium oxide is dispersed into the deuterated layer at a concentration of 1.3~at\% as a tracer for X-ray spectroscopic diagnostics.
To prevent direct laser interaction with the titanium-doped region, undoped polystyrene ($\mathrm{C_8H_8}$) layers, each $1~\mu \mathrm{m}$ thick, were placed on both sides of the film.
A crystal spectrometer was positioned at an angle of $28.9^\circ$ relative to the LFEX laser axis to measure the X-ray emission from the doped titanium and evaluate the electron temperature. The electron spectrometer was placed at an angle of $10.8^\circ$ from the laser axis to measure the energy distribution of fast electrons.

\section{Enhancement of Focused Intensity by High-Contrast Laser and Cone}
\label{sec3}

\begin{figure}[htbp]
  \centering
  \begin{minipage}[h]{\linewidth}
    \centering
    \includegraphics[width=\linewidth]{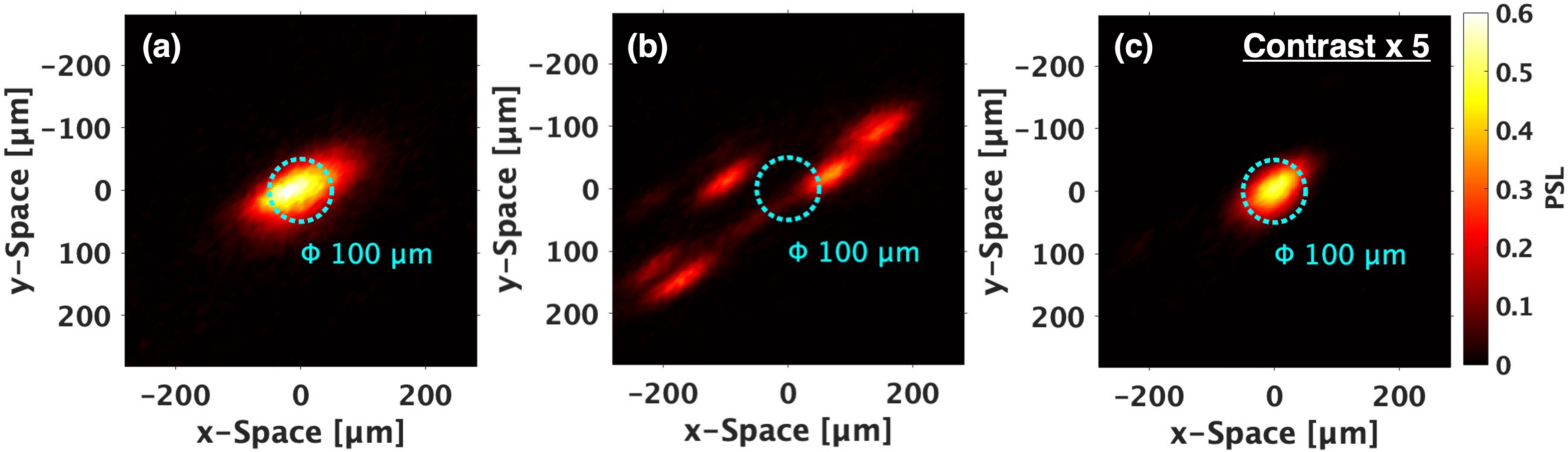}
  \end{minipage}
  \caption{
    X-ray images of the laser focal spot observed using the X-ray pinhole camera. 
    The two-dimensional images were projected along the target normal direction using a projective transformation.
    The dashed circle represents a region with a diameter of 100~{\textmu}m.
    (a) Low-contrast laser irradiation on the thin-film target. 
    (b) High-contrast laser irradiation on the thin-film target, configured with a multi-spot pattern. 
    (c) High-contrast multi-spot laser irradiation on a thin-film target with an attached cone.
  }
  \label{fig:Pinhole camera}
\end{figure}

Figure~\ref{fig:Pinhole camera} shows two-dimensional distributions of X-ray emission captured by the X-ray pinhole camera.  
The horizontal and vertical axes represent spatial position in units of {\textmu}m, and the color scale indicates photostimulated luminescence (PSL) values.  
A projective transformation was applied to convert the recorded images to a view along the target normal direction.  
The dashed circle in each panel marks a region with a diameter of 100~{\textmu}m.

Panel (a) shows the result of low-contrast laser irradiation onto the thin-film target, confirming that the laser focal spot was contained within a 100~{\textmu}m-diameter region.  
Panel (b) corresponds to high-contrast laser irradiation on the thin-film target. A multi-spot pattern was intentionally applied to evaluate the cone’s focusing effect.  
Panel (c) presents the case of high-contrast, multi-spot laser irradiation onto a thin-film target with an attached cone.  
It is evident that the laser components, which were initially distributed around the periphery, are redirected and concentrated at the cone tip.  
\begin{figure}[htbp]
  \centering
  \begin{minipage}{\textwidth}
    \centering
    \includegraphics[width=0.5\linewidth]{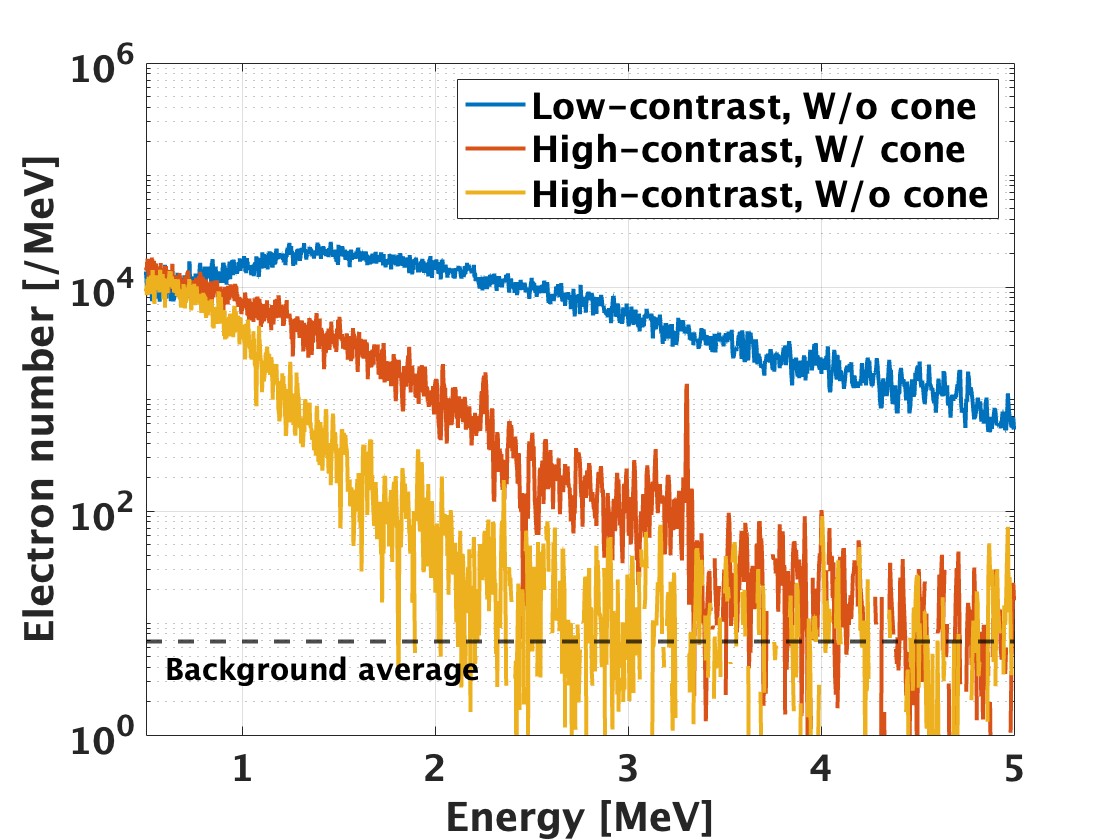}
  \end{minipage}
  \caption{
    Energy distributions of fast electrons measured by the electron spectrometer.  
    The horizontal axis indicates the electron energy [MeV], and the vertical axis represents the number of electrons per 1~MeV.  
    Blue line: Low-contrast laser (thin-film target);  
    Yellow line: High-contrast laser (thin-film target);  
    Red line: High-contrast laser (thin-film target with cone);  
    Black dashed line: Average background signal.
  }
  \label{fig:SlopeTemp}
\end{figure}

Figure~\ref{fig:SlopeTemp} shows the energy distributions of fast electrons measured with the electron spectrometer.  
The horizontal axis represents the electron energy in MeV, and the vertical axis indicates the number of electrons per 1~MeV.
The measured spectra were fitted using a Maxwell-Boltzmann distribution function,
\[
  f(E) = A \exp\left(-\frac{E}{T_{\mathrm{fast}}}\right),
\]
from which the electron slope temperature $T_{\mathrm{fast}}$ was extracted.  
The fitting uncertainty was evaluated using the standard error derived from the residuals between the experimental data and the fitted function.

For the flat target irradiated with a low-contrast laser, the slope temperature was measured to be $T_{\mathrm{fast}} = 972 \pm 6~\mathrm{keV}$.  
Using the empirical scaling between electron slope temperature and laser intensity derived from previous studies involving pre-plasma~\cite{pukhov1999particle},
\[
  T_{\mathrm{h}}~[\mathrm{MeV}] = 1.5 \sqrt{I},
\]
Here, $I$ is the laser intensity in units of $10^{18}~\mathrm{W/cm^2}$.
The corresponding laser intensity was estimated to be $0.4 \times 10^{18}~\mathrm{W/cm^2}$.

When using a high-contrast laser, the measured slope temperatures were $T_{\mathrm{fast}} = 125 \pm 11~\mathrm{keV}$ for the flat target without a cone, and $T_{\mathrm{fast}} = 412 \pm 7~\mathrm{keV}$ for the target with an attached cone.  
These values were further analyzed using the ponderomotive scaling law proposed by Wilks et al.~\cite{wilks1992absorption}:
\[
  T_{\mathrm{h}}~[\mathrm{MeV}] = 0.511 \left( \sqrt{1 + \frac{I \lambda^2}{1.37}} - 1 \right),
\]
which describes fast electron acceleration due to the ponderomotive force.  
This model is particularly applicable to high-contrast laser conditions, where long-scale pre-plasma formation is suppressed.
Here, $\lambda$ is the laser wavelength in micrometers.

Based on this scaling, the estimated laser intensity was $I = 0.7 \times 10^{18}~\mathrm{W/cm^2}$ for the high-contrast case without a cone, and $I = 2.8 \times 10^{18}~\mathrm{W/cm^2}$ for the case with a cone.  
This indicates that the presence of the cone resulted in approximately a fourfold enhancement of laser intensity at the focal region.

These results suggest that the cone structure effectively improves laser focusing, thereby enhancing the intensity of the laser light delivered to the target.

\section{Evaluation of Plasma Electron Temperature}
\label{sec4}

The plasma electron temperature was evaluated by analyzing the X-ray emission spectra from titanium doped into the target material.  
The experimental titanium spectra were compared with those calculated using collisional-radiative simulations~\cite{matsuo2020_petapascal}.  
The simulations were performed using the PrismSPECT code~\cite{macfarlane2003simulation}, which calculates X-ray spectra based on collisional-radiative atomic models.

In this study, atomic data were generated using the Flexible Atomic Code (FAC)~\cite{Gu2008}.  
The ionization state distribution of the plasma was obtained by solving the rate equations, which account for major atomic processes including collisional ionization, excitation and de-excitation, radiative and three-body recombination, two-electron recombination, autoionization, photoionization, and photoexcitation.

The plasma thickness was set to 1~{\textmu}m at solid density, and steady-state conditions were assumed for the calculations.
The target composition is identical to the experimental material and given as follows:  
H: 11~at\%, D: 39~at\%, C: 45~at\%, Ti: 1.3~at\%, Si: 0.52~at\%, and O: 3.6~at\%.  

To account for inner-shell ionization caused by fast electrons, a two-electron-component model was adopted.  
In this model, the fast electron component was treated separately from the bulk electron distribution.  
The energy distribution of fast electrons was determined based on the slope temperatures obtained from the electron spectrometer.  
Both the bulk electron temperature and the fast electron mixing ratio were used as input parameters for the spectral calculations.

The similarity between the simulated and experimental spectra was quantified using the Pearson correlation coefficient, defined as:
\[
  \rho(A, B) = \frac{1}{N - 1} \sum_{i = 1}^{N} \left( \frac{A_i - \mu_A}{\sigma_A} \right) \left( \frac{B_i - \mu_B}{\sigma_B} \right),
\]
where $\mu_A$ and $\mu_B$ are the means of $A$ and $B$, and $\sigma_A$ and $\sigma_B$ are their standard deviations, respectively.

The correlation was evaluated over the photon energy range of 5.3 - 6.0~keV, where the effect of self-absorption is minimal.  
This range includes the ${\rm He}\beta$, ${\rm Ly}\beta$, and ${\rm He}\gamma$ transitions of titanium ions.

The electron temperature and the fast electron mixing ratio were estimated by identifying the conditions that maximized the correlation coefficient.  
The uncertainty in the estimated temperature was determined from the range corresponding to the 95\% confidence interval of the correlation coefficient.

\begin{figure}[htbp]
  \centering
  \begin{minipage}{\textwidth}
    \centering
    \includegraphics[width=\linewidth]{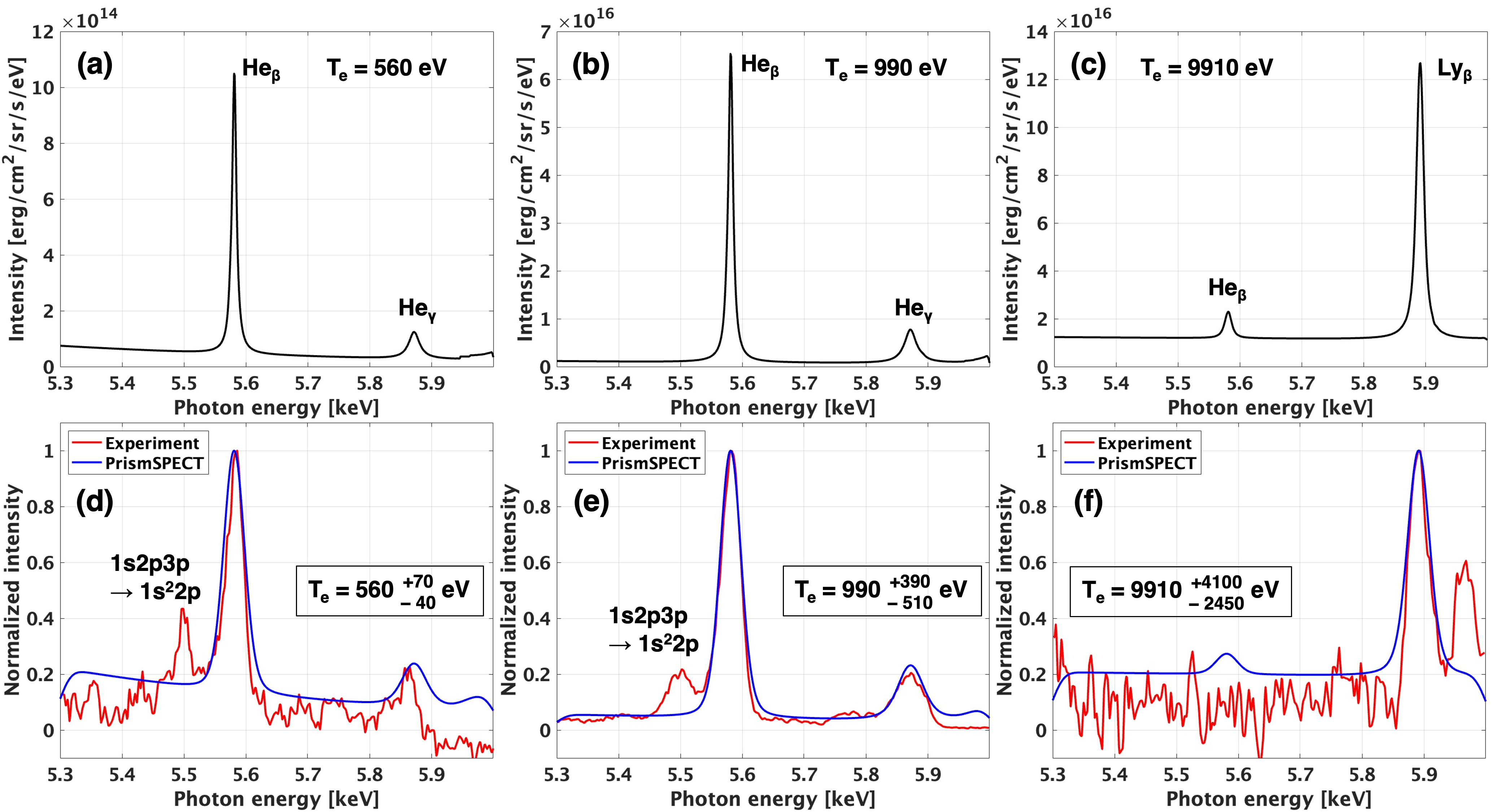}
  \end{minipage}
  \caption{
    Top panels (a) - (c): Simulated titanium spectra calculated using PrismSPECT.  
    Bottom panels (d) - (f): Comparison between Gaussian-convolved simulated spectra (FWHM = 15~eV, blue) and experimental spectra (red).  
    (d) Low-contrast laser irradiation.  
    (e) High-contrast laser irradiation.  
    (f) High-contrast laser irradiation with a cone-attached target.
  }
  \label{fig:X-Ray spectra}
\end{figure}

Figure~\ref{fig:X-Ray spectra}(a) - (c) show titanium spectra calculated using PrismSPECT at electron temperatures of $T_{\mathrm{e}} =$ 560~eV, 990~eV, and 9910~eV, respectively.  
Figures~\ref{fig:X-Ray spectra}(d) - (f) present comparisons between the Gaussian-convolved simulated spectra (FWHM = 15~eV) and the corresponding experimental spectra.

The peaks observed around 5.5~keV in panels (d) and (e) are attributed to the $1s\,2p\,3p \rightarrow 1s^2\,2p$ transitions of Ti$^{19+}$ ions, likely originating from doubly excited states.  
However, these features are not reproduced in the current simulations.  
Additionally, a spectral structure observed near 5.95~keV in panel (f) could not be identified, as no corresponding transition line is listed in the NIST atomic database~\cite{ralchenko2005nist}.

X-ray spectral analysis revealed that the use of a high-contrast laser increased the plasma electron temperature to approximately 1~keV, as seen in Figures~\ref{fig:X-Ray spectra}(d) and (e).  
Furthermore, when a cone target was used, the electron temperature increased substantially, reaching approximately 9~keV, as shown in Figure~\ref{fig:X-Ray spectra}(f).  
This value is roughly 17.5 times higher than that obtained with low-contrast laser irradiation of a flat target.
These results indicate that the combination of high-contrast laser irradiation and cone-target geometry significantly enhances plasma heating efficiency.

\section{Conclusion \& future perspective}
\label{sec5}

We demonstrated that combining a high-intensity, high-contrast laser with a cone-attached target significantly enhances plasma heating efficiency. By irradiating a titanium-doped deuterated polystyrene target equipped with a gold cone, we observed a significant increase in the electron temperature of a heated target, indicating a synergistic interaction between the high-contrast laser and the cone structure.

The electron temperature of the heated plasma was estimated to be approximately 9 keV by comparing measured resonant X-ray spectra with those calculated using the atomic physics code Prism-SPECT. This enhanced heating is attributed to the reflection of the high-contrast laser from the inner surface of the cone, which increases the laser intensity at the cone tip. Supporting this, the energy spectrum of the fast electrons shifted toward higher energies.

The electron temperature was estimated under the assumption of a spatially uniform, steady-state plasma using Prism-SPECT. However, a more detailed evaluation that accounts for spatial gradients in temperature and density requires radiation transport modeling with SPECT3D~\cite{macfarlane2007spect3d}. At 9 keV, the plasma becomes dominated by fully ionized ions. To accurately model atomic processes, including recombination in transient, fully ionized plasmas, time-resolved spectral measurements and time-dependent simulations are necessary.
To evaluate the temperature gradient along the laser propagation axis, we plan to employ multilayered targets embedded with multiple tracer elements at different depths. This measurement is essential to determine whether collisional heating by fast electrons or thermal conduction from laser-heated regions is the dominant heating mechanism~\cite{kemp2006collisional}.
Furthermore, to fully leverage the focusing effect of the cone structure, we will explore the application of high-contrast lasers in combination with compound parabolic concentrators (CPCs)~\cite{macphee2020_CPC, williams2021order}.

\section{Acknowledgements}
\label{sec6}
The authors thank the technical support staff at The University of Osaka for assistance with laser operation, target fabrication, plasma diagnostics, and computer simulation. 
This work was partly achieved through the use of large-scale computer systems at D3 Center at The University of Osaka.
This work was supported by the Joint Usage/Research Center program of the Institute of Laser Engineering (ILE) at The University of Osaka (2023A1-017FUJIOKA); the Collaboration Research Program between the National Institute for Fusion Science and ILE (2020NIFS12KUGK057) and Grants-in-Aid for Scientific Research (Nos. 25K17369, 23K03360, 22H00118, 22H01205, 22H01206, 22K03567, 21H04454, 20H00140, 20H01886, 17K05728, and 16H02245), "Power Laser DX Platform" as research equipment shared in the Ministry of Education, Culture, Sports, Science and Technology Project for promoting public utilization of advanced research infrastructure (Program for advanced research equipment platforms, grant number JPMXS0450300021); the Japan Society for the Promotion of Science Core-to-Core Program  (grant number JPJSCCA20230003); 
R. T. and J. D. are partially supported by the QLEAR fellowship program of Osaka University.










\bibliographystyle{elsarticle-num.bst}
\bibliography{MyRef}
\end{document}